\documentclass[aps,aps,ams,amsmath,showpacs,floats]{revtex4}

\usepackage{pifont,multirow,graphicx,    %color,colortbl,multicol,
makeidx,rotating,fancyhdr,fancybox,calc,amsmath,longtable,
amsfonts,amssymb,amsthm,latexsym,epsfig,epstopdf} 

\usepackage[english]{babel}

\usepackage[latin1]{inputenc}
  {\left\lbrace\begin{array}{@{}l@{}}}%
  {\end{array}\right.}
  
\newcommand{\nc}{\newcommand}

\nc{\be}{\begin{equation}}
\nc{\ee}{\end{equation}}
\nc{\bea}{\begin{eqnarray}}
\nc{\eea}{\end{eqnarray}}
\nc{\ba}{\begin{array}}
\nc{\ea}{\end{array}}
\nc{\ov}{\overline}
\nc{\wt}{\widetilde}
\nc{\linea}{}
\nc{\vare}{\varepsilon} \nc{\dis}{\displaystyle}
\begin{document}

\title{High sensitivity measurements of thermal properties of textile fabrics}
\author{D.\;Romeli}
\email{romelid@libero.it}% 
\affiliation{Universit\`a di Bergamo, Facolt\`a di Ingegneria, \\ viale Marconi 5,
I-24044 Dalmine (BG), Italy}

\author{G.\;Barigozzi}
\email{giovanna.barigozzi@unibg.it}%
\affiliation{Universit\`a di Bergamo, Dipartimento di Ingegneria, \\ viale Marconi 5,
I-24044 Dalmine (BG), Italy }

\author{S.\;Esposito}
\email{Salvatore.Esposito@na.infn.it}%
\affiliation{Istituto Nazionale di Fisica Nucleare, Sezione 
di Napoli, Complesso Universitario di Monte S. Angelo, via Cinthia, 
I-80126 Naples, Italy}

\author{G.\;Rosace}
\email{giuseppe.rosace@unibg.it}%
\affiliation{Universit\`a di Bergamo, Dipartimento di Ingegneria, \\ viale Marconi 5,
I-24044 Dalmine (BG), Italy}

\author{G.\;Salesi}
\email{salesi@unibg.it}%
\affiliation{\mbox{Universit\`a di Bergamo, Facolt\`a di Ingegneria, viale Marconi 5,
I-24044 Dalmine (BG), Italy} \\ %}}
%\affiliation{
Istituto Nazionale di Fisica Nucleare, Sezione di Milano, via
Celoria 16, I-20133 Milan, Italy }

\begin{abstract}

\noindent A new testing apparatus is proposed to measure the thermal properties of fabrics made from polymeric materials. The calibration of the apparatus and the data acquisition procedure are considered in detail in order to measure thermal conductivity, resistance, absorption and diffusivity constants of the tested fabric samples. Differences between dry and wet fabrics have been carefully detected and analyzed. We have developed a new measurement protocol, the {\it ThermoTex} protocol, which agrees with the UNI EN 31092 standard and entails an accurate quantification of the experimental errors according to a standard statistical analysis, thus allowing a rigorous investigation of the physical behavior of the phenomena involved. As a consequence, our machinery exhibits great potentialities for optimizing the thermal comfort of fabrics, according to the market demand, thanks to the possible development of a predictive phenomenological theory of the effects involved.
\end{abstract}

\maketitle

\section{Introduction}

\noindent Thermal properties of textile fabrics are of great interest and relevance for textile researchers \cite{i}-\cite{iv}, since they determine several among the major characteristics related to the wearer's overall comfort perception \cite{v}, having significant effects on the skin mean temperature too \cite{vi}. The heat transfer through clothes takes place by means of three distinct processes: conduction, convection and radiation. In the conduction mechanism heat propagates through short range interactions of molecules and/or electrons, while convection involves heat transfer by means of the combined mechanisms of fluid mixing and conduction; instead heat radiates away in the form of electromagnetic waves, mainly in the infrared (IR) region \cite{vii}. For what concerns here, it is generally accepted that heat transfer by conduction is more significant than other mechanisms. 

For a textile material composed of fibers and entrapped air, the thermal conductivity is the result of a combination of two effects mediated by the fiber polymer and air thermal conductivity. Within the fibrous insulation, heat transfer  mainly involves conduction through the solid material of the fibers, convection through the intervening air and thermal radiation, but a lot of literature has shown that radiative heat transfer could lead to a significant contribution to the total heat transfer within these highly porous media \cite{viii}. Moreover, textile surface treatments and finishing processes strongly influence the textile thermal insulating properties, and this is particularly true for those related to the IR thermal radiation \cite{v}. 

In the literature \cite{iii,x,xii,xiii} many papers are devoted to thermal insulation, comfort properties of clothing and to the related experimental techniques and measurement methods. At the same time, several existing standard procedures and testing methods have been developed in order to specify fabric IR properties \cite{xiv,xv}. Basically, some of the proposed approaches are based on the establishment of a steady-state thermal conductivity regime where an electric heater provides a temperature field in a given sample \cite{xvi}. In addition, other experimental devices have been proposed in recent years to test the thermal comfort of textile fabrics, like the ``hot disk'' \cite{xvii} and the FRMT \cite{vii} ones, this machinery being successfully applied \cite{xix,xx} to the characterization of several kinds of textile fabrics, including the effect of different fabric covering factors and finishing agents. Two such devices are commercially available: the {\it Thermolabo KES-FB7} system developed by Kawabata \cite{xxi} and the {\it Alambeta} apparatus built by Hes and Dolezal \cite{xxii,xxiii}, the last one allowing the measurement of the textile thermal contact properties which enter into the definition of the so-called ``warm/cool feeling''.
 
The measurement protocol of thermal and transpiration properties is coded in the European Standard UNI EN 31092 \cite{xxiv}. This code is based on the use of a steady-state device -- the so-called ``Skin Model'' --simulating the amounts of heat and humidity exchanged between the human body and the external environment through the clothes worn. However, the methods employed have the disadvantage of measuring the fabric surface temperature in a single point, and then assuming a uniform temperature distribution over the textile surface, which is certainly not the case for fabrics characterized by low covering factors showing highly variable temperature values over their structure. In the present paper we relax such a limiting assumption and present several important results coming from a careful experimental investigation aimed at studying the thermal characteristics of different fabrics of variable yarn configuration. The results obtained will form the basis of a future theoretical analysis which will be able to predict the fabric thermal behavior, taking into account different influencing parameters, such as the morphology of the component fibers, the yarn structure and properties, the physical and structural characteristics of fabrics and the effect of finishing agents \cite{xxv}. The experimental apparatus developed here is similar to the hot disk mentioned above, but takes advantage of the use of an IR thermocamera to measure the whole fabric surface temperature distribution.

\section{Experiments}

\subsection{Materials}

\noindent In order to verify the accuracy of the methodology proposed, we first tested insulating materials of known thermal properties, like neoprene and polystyrene, and then considered cotton samples, both dry and wet: the main characteristics of textile samples used in this research are reported in Table \ref{tab1}. The fabrics were washed in a $2\%$ non-ionic detergent (BERDET WF, wetting agent, kindly supplied by Europizzi, Urgnano, Italy) at $pH=7$ and temperature of $40 \, ^{\rm o}$C for 20 minutes, and then rinsed several times with deionized water, dried and put into drier for storage. The cleaned samples were conditioned under standard atmospheric pressure at $65 \pm 4 \%$ relative humidity and at a temperature of $20 \pm 2 \, ^{\rm o}$C for at least 24h before any of our experiments. 

\begin{table}%[!h]
\centering\footnotesize
\begin{tabular}{|c|c|c|c|c|c|}
\hline \textbf{Code} & \textbf{Material} & \textbf{Type} & \textbf{Thickness} [$10^{-5}$m] & \textbf{Mass per unit area} [g/m$^2$] & \textbf{Density} [kg/m$^3$] \\
\hline C1     &  Polyester  &          & $700   \pm 100$     &  & \\
\hline C2     &  Neoprene  &       & $1200 \pm 100$     &  & \\
\hline C3     &  Polystyrene  &    & $1500 \pm 100$     &  & \\
\hline T11A &  Cotton  &  satin  & $31         \pm 1$ & $160.4 \pm 0.8$ & $520 \pm 20$\\
\hline T15A &  Cotton  &  satin  & $38         \pm 3$ & $179.6 \pm 0.9$ & $470 \pm 40$\\
\hline T20A &  Cotton  &  satin  & $37         \pm 1$ & $213       \pm 1$ & $580 \pm 20$ \\
\hline T25A &  Cotton  &  satin  & $41         \pm 1$ & $242       \pm 1$ & $590 \pm 20$ \\
\hline T11B &  Cotton  &  plain  & $30         \pm 1$ & $166.1 \pm 0.8$ & $550 \pm 20$ \\
\hline T15B &  Cotton  &  plain  & $31         \pm 1$ & $177.1 \pm 0.9$ & $570 \pm 20$ \\
\hline T20B &  Cotton  &  plain  & $32         \pm 1$ & $199       \pm 1$ & $620 \pm 10$ \\
\hline T25B &  Cotton  &  plain  & $35         \pm 3$ & $223       \pm 1$ & $640 \pm 50$ \\
\hline
\end{tabular}
\caption{{Some properties of the materials employed in the present study}}
\label{tab1}
\end{table}

\subsection{Test methods}

\noindent As well known, any thermal property of a given fabric depends on the thermal conductivity $\lambda$, the latter being a function of the thermal power $\dot{Q}$ transmitted through the testing sample, of its surface $A$ and thickness $s$, and of the temperature difference $\Delta T$ across its thickness:
\begin{equation}
\dis{\lambda = \frac{\dot{Q}\cdot s}{A\cdot \Delta T}} \, .
\label{eq1}
\end{equation}
The thermal comfort properties like thermal resistance $R$, absorption $b$ and diffusion $a$ are derived from the measured thermal conductivity $\lambda$ and are computed as follows:
\begin{eqnarray}
& & R = \frac{s}{\lambda} \, , \label{eq2} \\
& & b = \sqrt{\lambda \rho c} \, \label{eq3} \\
& & a = \frac{\lambda}{\rho c} \, ,  \label{eq4}
\end{eqnarray}
where $\rho$ [kg/m$^3$] is the density of the sample (calculated by dividing the surface density by thickness) and $c$ [J/(kgK)] its specific heat. The thermal resistance $R$ is associated with the fabric structure, and is a very important parameter in ruling thermal insulation. The thermal absorption $b$ is, instead, a surface property that can be related to the warm/cool feeling: fabrics with a low value of the thermal absorption will give a warm feeling. Finally, the thermal diffusion $a$ is a property describing the heat flow through the fabric. 

\begin{figure}%[!ht]
\centering 
\includegraphics[scale=0.6]{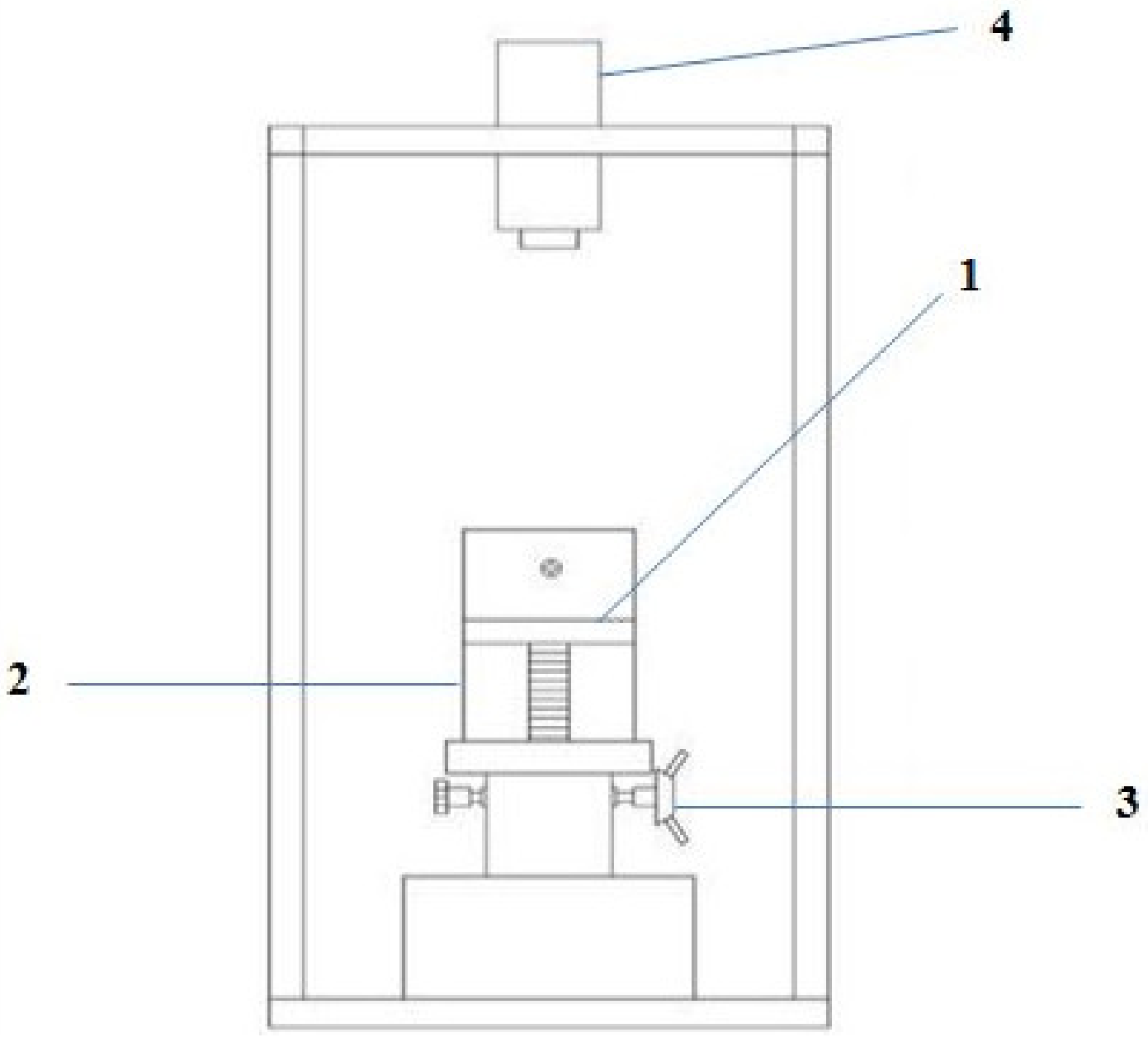}
\includegraphics[scale=0.7]{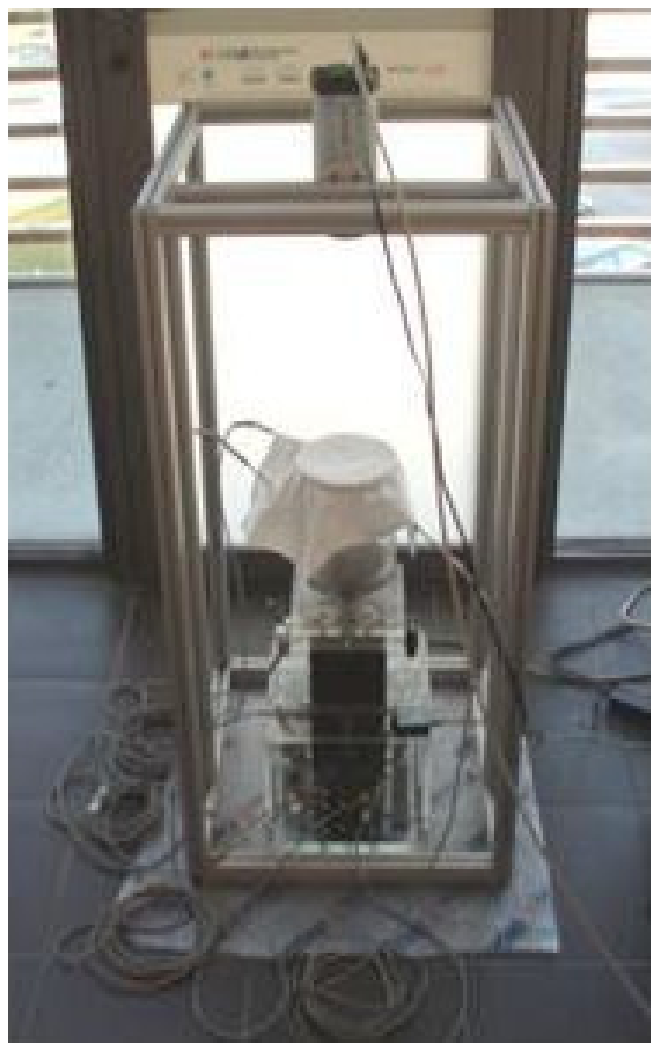}
\caption{A view of the TheCotex apparatus: 1) hot plate; 2) plexiglas cylinder; 3) control mechanism for raising/lowering the hot plate; 4) IR thermocamera.}
\label{fig1}
\linea
\end{figure}

In order to determine such properties, a new test method was developed, based on the principle that the conductivity $\lambda$ is measured through the ratio between a known thermal power and the temperature difference in a system where conduction prevails. The experimental setup was thus designed to provide measurement of the cooling rate of a solid body heated to a fixed temperature and insulated from the ambient medium by the fabric. The schematic diagram of the experimental apparatus, named {\it TheCotex}, is shown in Fig.\,\ref{fig1}. It consists in a heating device, i.e. an hot plate, that can be moved inside an externally insulated plexiglas cylinder, on the top of which the fabric sample is installed, thus separating the hot cylindrical cavity from the ambient air. An automatic control system allows to maintain the hot plate temperature at the desired working point with an accuracy of  $\pm 0.1 \%$ and a stability of $\pm 0.1 \, ^{\rm o}$ C. In order to minimize heat losses, the hot plate was directly in contact with the textile sample, so that the plate temperature, measured by means of calibrated T-type thermocouples, corresponds just to the bottom temperature of the sample. An IR thermocamera (FLIR ThermoVisionTM A40 with a 64/150 mm close-up lens) is used to capture the temperature distribution over the external surface of the textile sample but, in addition to it, other T-type thermocouples are also glued to the external surface of the sample for calibration purposes. The heat flux across the sample is finally calculated starting from the electric power dissipated into the hot plate. 

A calorimeter was, then, manufactured to characterize the TheCoTex operation: it allows to evaluate the actual thermal power $Q$ from the hot plate to the fabric sample. Indeed, due to heat losses through the ambient air and the plate support, it could differ from the measured electrical power $P_{\rm el} = VI$, given by the input voltage $V$ and the hot plate operating current $I$. This procedure leads to the definition of a effective {\it correction curve}, in order to estimate the actual thermal power from the measured supplied electrical power. 

The following protocol has been adopted for the measurement calorimetric procedure:
\begin{enumerate}
\item fix the hot plate input voltage;
\item measure the actual thermal power every ten minutes (for at least two hours);
\item repeat point 2 (at least one more time), taking care that the hot plate is turned off between a given run and the following one, until the complete cooling;
\item repeat points 2 and 3 for at least three different input voltages (for example: 2, 3, 4 V).
\end{enumerate}

\begin{figure}%[!ht]
\centering 
\begin{tabular}{ccc}
\includegraphics[scale=0.88]{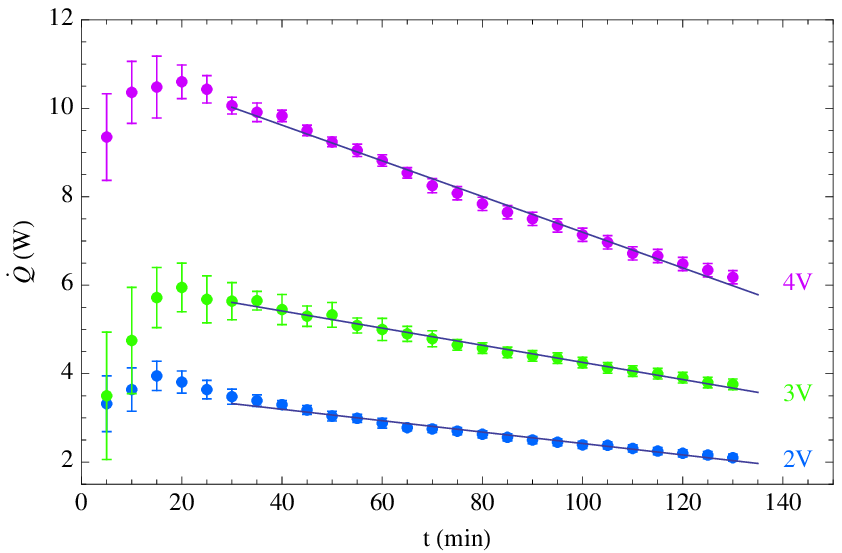} a) & \hspace{1.5cm} & \includegraphics[scale=0.35]{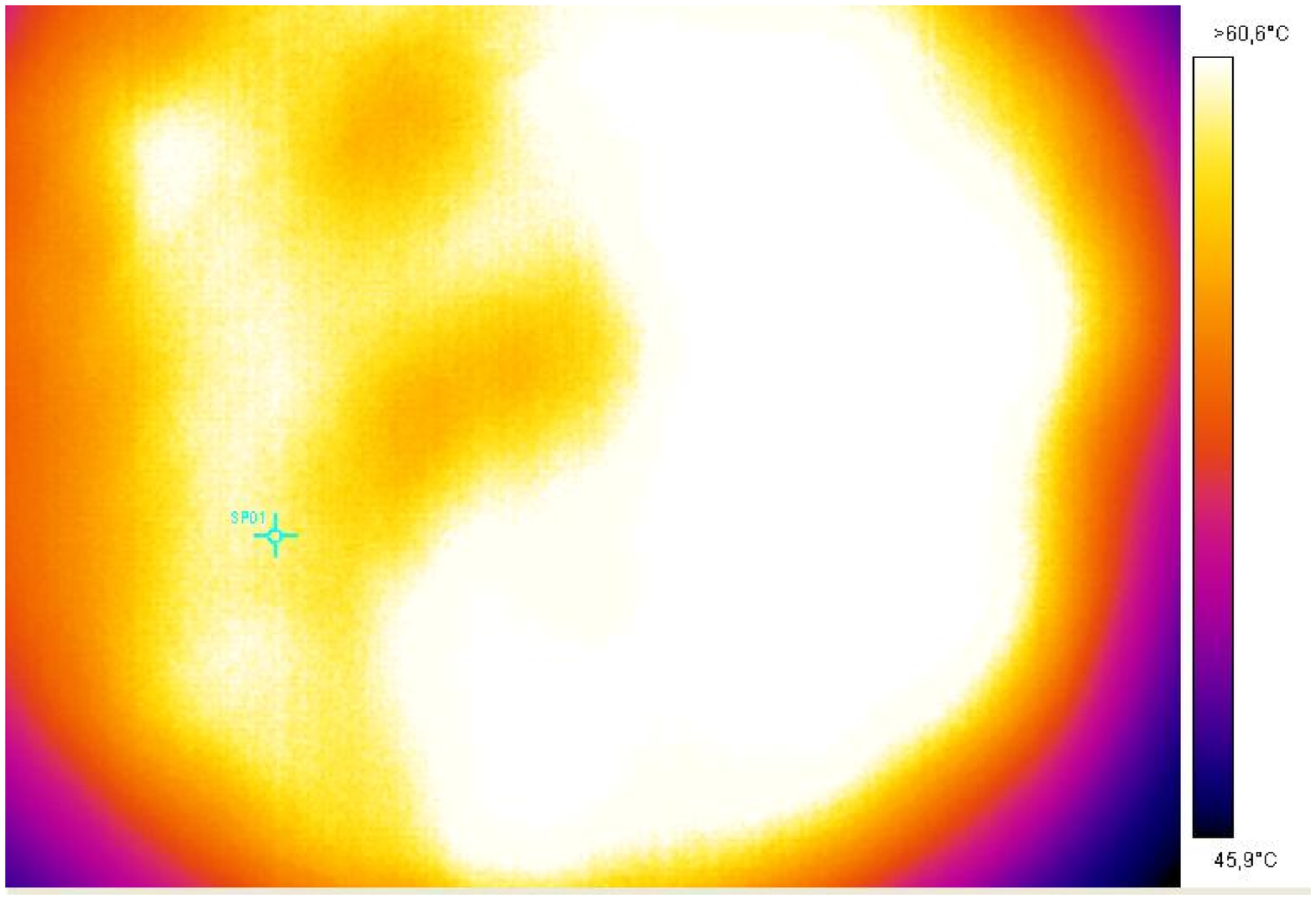} b)
\end{tabular}
\caption{a) Time evolution of the actual thermal power of the hot plate, for different input voltages. b) A sample thermography obtained with the IR thermocamera, showing the reaching of temperature homogeneity of the hot plate.}
\label{fig2}
\linea
\end{figure}

Figure \ref{fig2}a shows the actual thermal power versus time transmitted by the hot plate, for different values of the input voltage: the different curves exhibit a similar trend, with a maximum around the minute 20 (after 
the hot plate was switched on), followed by a slow decrease of the power.

\section{Results and discussion}

\subsection{Calibration}

\begin{figure}%[!ht]
\centering 
\includegraphics[scale=0.72]{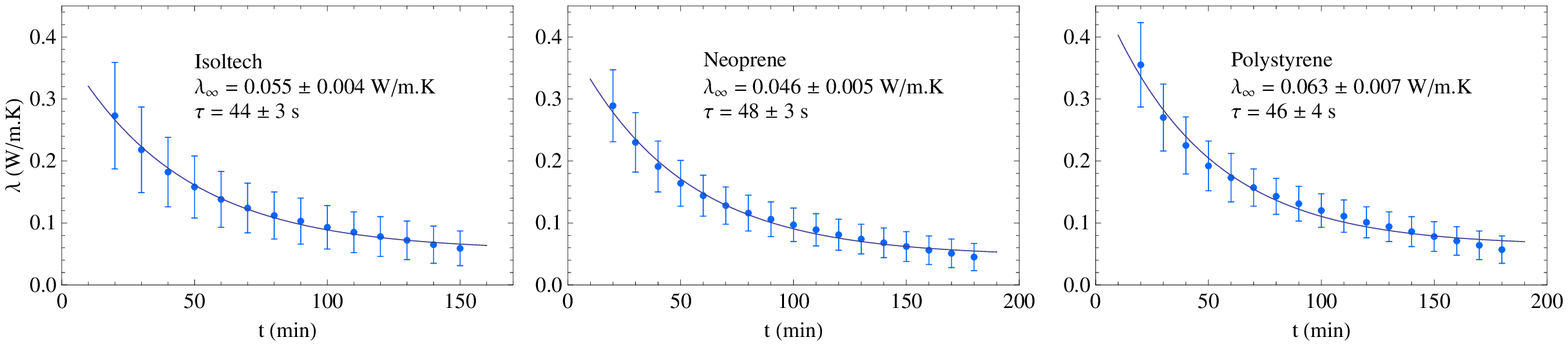}
\caption{Calibration measurements: experimental data on the thermal conductivity of polyester (a), neoprene (b) and polystyrene (c) obtained with the TheCotex device (by using thermocouples). The fitting curves correspond to the exponential behavior in Eq.\,(\ref{eq5}).}
\label{fig3}
\linea
\end{figure}

\noindent The efficacy of the calibration model described above has been tested by making measurements of the thermal transmittance of several insulating materials (polyester, neoprene and polystyrene). The quantity $\lambda$ is obtained from Eq.\,(\ref{eq1}), where $Q$ is the power emitted by the hot plate at a given time and  voltage, while $s$ and $A$ are the thickness and the surface of the sample (in contact with the hot plate), respectively, and $\Delta T$ is the temperature difference between the upper and the lower surface of the sample. From our data, collected at different times, we see (Fig.\,\ref{fig3}) that the behavior of $\lambda$ is not stationary, but it decreases with time. The fitting curve we adopted is a simple exponential one, tending to a finite value for the time $t$ tending to infinity:
\begin{equation}
\lambda = \lambda_\infty + \lambda_0 {\rm e}^{- t/\tau} \, .
\label{eq5}
\end{equation}
The asymptotic value $\lambda_\infty$ is, then, directly deduced from standard regression of this law to our experimental data: it represents the regime thermal conductivity in stationary conditions. The above exponential model gives, however, also important informations about the thermal power transmitted by the hot plate. As a matter of fact, for {almost} any of the materials we investigated (and in any experimental condition), we obtain a time constant $\tau$ equal to about 40 minutes; {it is mainly independent of the material under study, while mainly depends on the hot played employed.}
%this indicating that it does not depend on the material under study, but rather mainly on the hot plate employed. 
The thermographic images (Fig.\,\ref{fig2}b), taken during the previous measurements, also indicate the existence of a typical critical time required by the device in order to reach a temperature homogeneity in any point of the hot plate. As expected, after this critical time, $\lambda$ reduces to the asymptotic value $\lambda_\infty$.

Such results thus confirm the validity of our calibration model of the heat source, and this is just the starting point for the settlement of the ThermoTex protocol, defined by the following measurements.

\subsection{Measurements I}

\noindent After the calibration of our machinery, the protocol has been applied to test cotton fabrics, obtaining their thermal conductivity and resistance, as well as their diffusivity and absorption constants. With the focus of a comparative analysis for the thermal behavior of a given sample, such properties have been analyzed as a function of several geometric, physical and tissue parameters, finding relevant differences among the different samples due to the sample geometry and to the presence of interstices full of air. The fabrics used in the present analysis were eight cotton samples, divided into two groups according to their warp/weft structure: with type $A$ we denote {\it satin} fabrics, while type B refers to {\it plain} fabrics.

\begin{figure}%[!ht]
\centering 
\begin{tabular}{cc}
\includegraphics[scale=0.8]{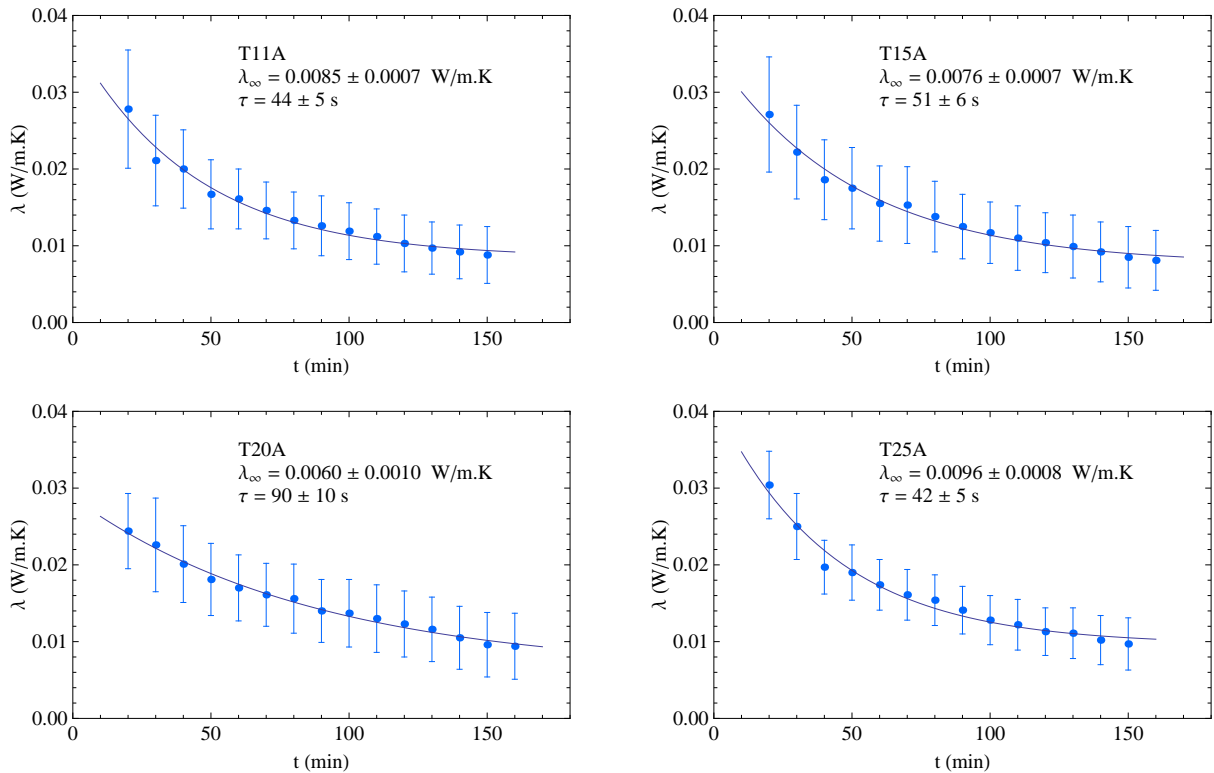}
\end{tabular}
\caption{Time evolution of the thermal conductivity of type $A$ fabrics.}
\label{fig4}
\linea
\end{figure}

Firstly, we measured thermal conductivities (as described above), whose values enter the evaluation of thermal resistance (see Fig.s\,\ref{fig4}). In order to check the results here obtained, % the samples were sent to an external laboratory \footnote{Centro Tessile Cotoniero e Abbigliamento S.p.A., Busto Arsizio (VA), Italy}, where 
the thermal resistance was measured also by means of the {\it Skin Model} method \cite{Borsi}. Our values, compared with those obtained from the Skin Model method, are reported in Table \ref{tab2}. From this, it is evident that almost all the different values agree each other within the experimental error bars we measured.

\begin{table}%[!h]
\centering
\begin{tabular}{|c|c|c|c|}
\hline \textbf{Fabric} & \textbf{ThermoTex} & \textbf{Skin Model} \\
\hline T11A &    $0.036 \pm 0.005$ & $0.037$ \\
\hline T15A &	$0.050 \pm 0.008$ & $0.034$ \\
\hline T20A &	$0.062 \pm 0.012$ & $0.035$ \\
\hline T25A &	$0.043 \pm 0.005$ & $0.031$ \\
\hline T11B &	$0.033 \pm 0.007$ & $0.022$ \\
\hline T15B &	$0.032 \pm 0.007$ & $0.024$ \\
\hline T20B &	$0.023 \pm 0.006$ & $0.020$ \\
\hline T25B &	$0.029 \pm 0.010$ & $0.020$ \\
\hline
\end{tabular}
\caption{Thermal resistance (measured in \rm{m$^2$K/W}) of different cotton samples as measured with the  ThermoTex protocol and as obtained with the Skin Model method}
\label{tab2}
\end{table}

\begin{figure}%[!ht]
\centering 
\begin{tabular}{cc}
\includegraphics[scale=0.55]{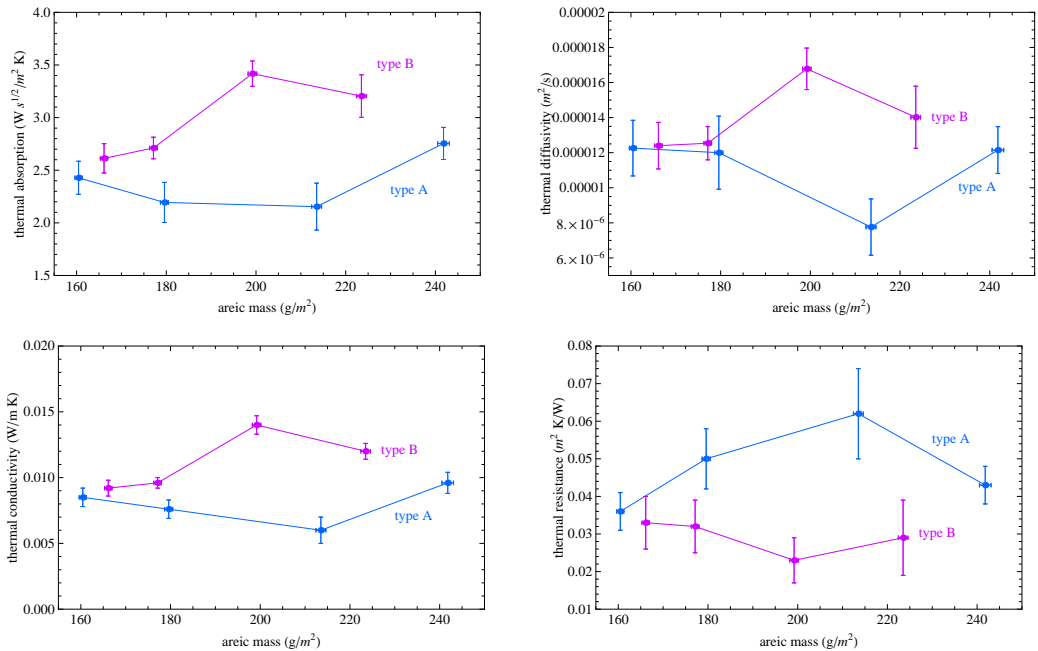}
\end{tabular}
\caption{Thermal absorption, diffusivity, conductivity and resistance of type $A$, $B$ samples versus the mass per unit area of the fabric.}
\label{fig5}
\linea
\end{figure}

A joined analysis of Tables \ref{tab1} and \ref{tab2} gives some useful information about the dependence of thermal resistance on the thickness, mass per unit area and linear density of the samples. Of some relevance is the ratio between the mass per unit area and the average value of the linear density, giving information on the trend of the weaving: for low values of this ratio, more space is left between the weft and the warp, while, for higher values, such space is much reduced. 
%The effective influence of the air in the interstices on the thermal response of the fabric can, then, be evaluated by comparing the dependence of the thermal resistance on the mass per unit area {(see Fig.\,\ref{fig5})} with that on the linear density. Indeed, while the value of the mass per unit area gives some physical information about the fabric, that of the density is especially related to geometric properties of the fabric, and the apparent different behavior coming from a great number of data picks out the important role played by the presence of air within the fabric. 
When considering also the other relevant properties, including absorption and diffusivity, interesting results are those presented in Fig.s\,\ref{fig5}. From these graphs we note a sharp difference of thermal conductivity and resistance between type A and type B fabrics, as well as a non-monotonic behavior of those quantities as a function of the mass per unit area. This might be correlated to the different warp/weft structure of a given fabric, and then to the different surface density and geometric area of the interstices (cfr. the next subsection about the influence of the air retention on the thermal properties of the fabrics). Particularly intriguing is the apparent existence of a {\it minimum} in some graphs, pointing out some extremal conditions in the thermal response of the fabric.

\subsection{Measurements II}

\begin{figure}%[!ht]
\centering 
\begin{tabular}{cc}
\includegraphics[scale=0.65]{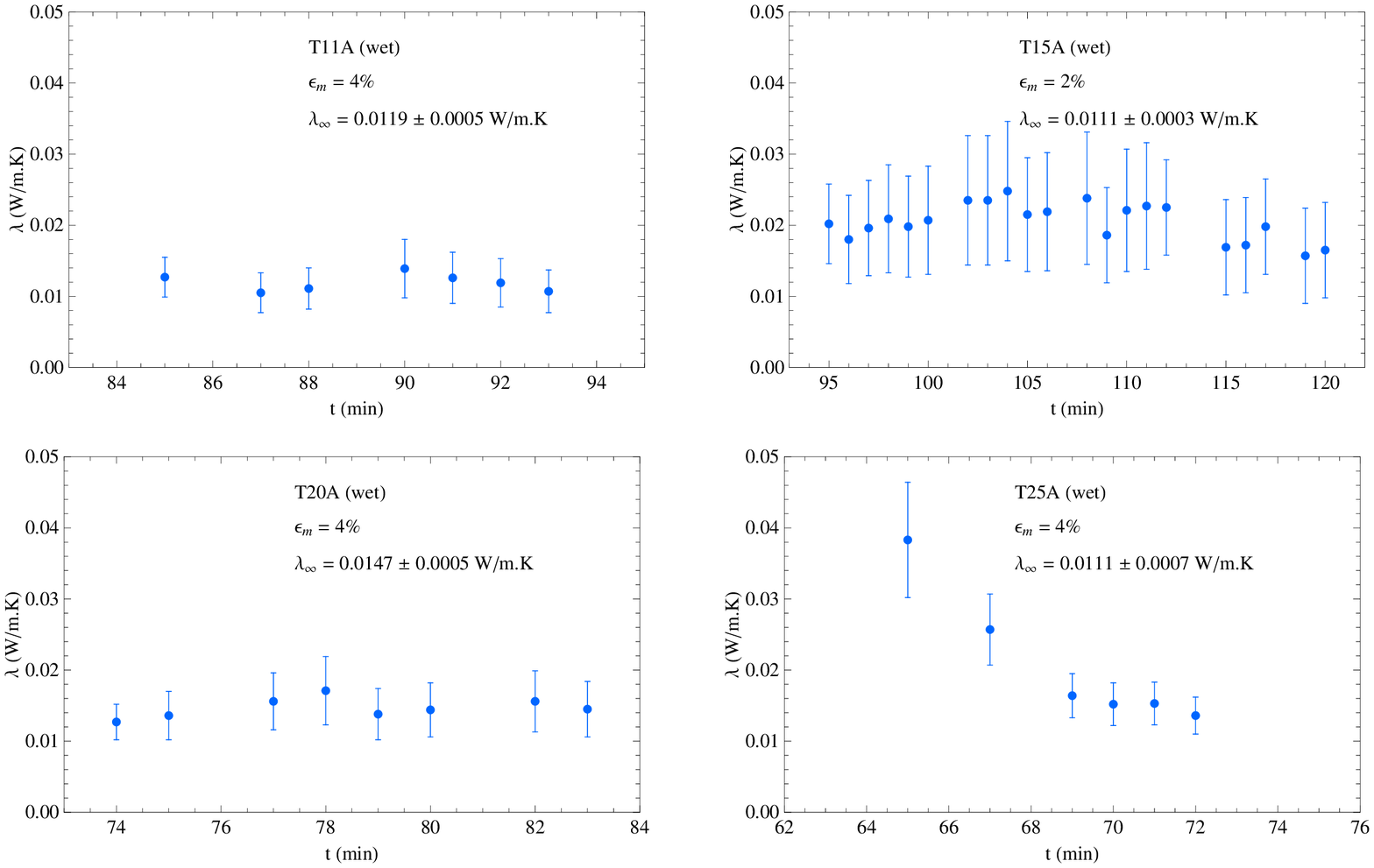} \\
\includegraphics[scale=0.65]{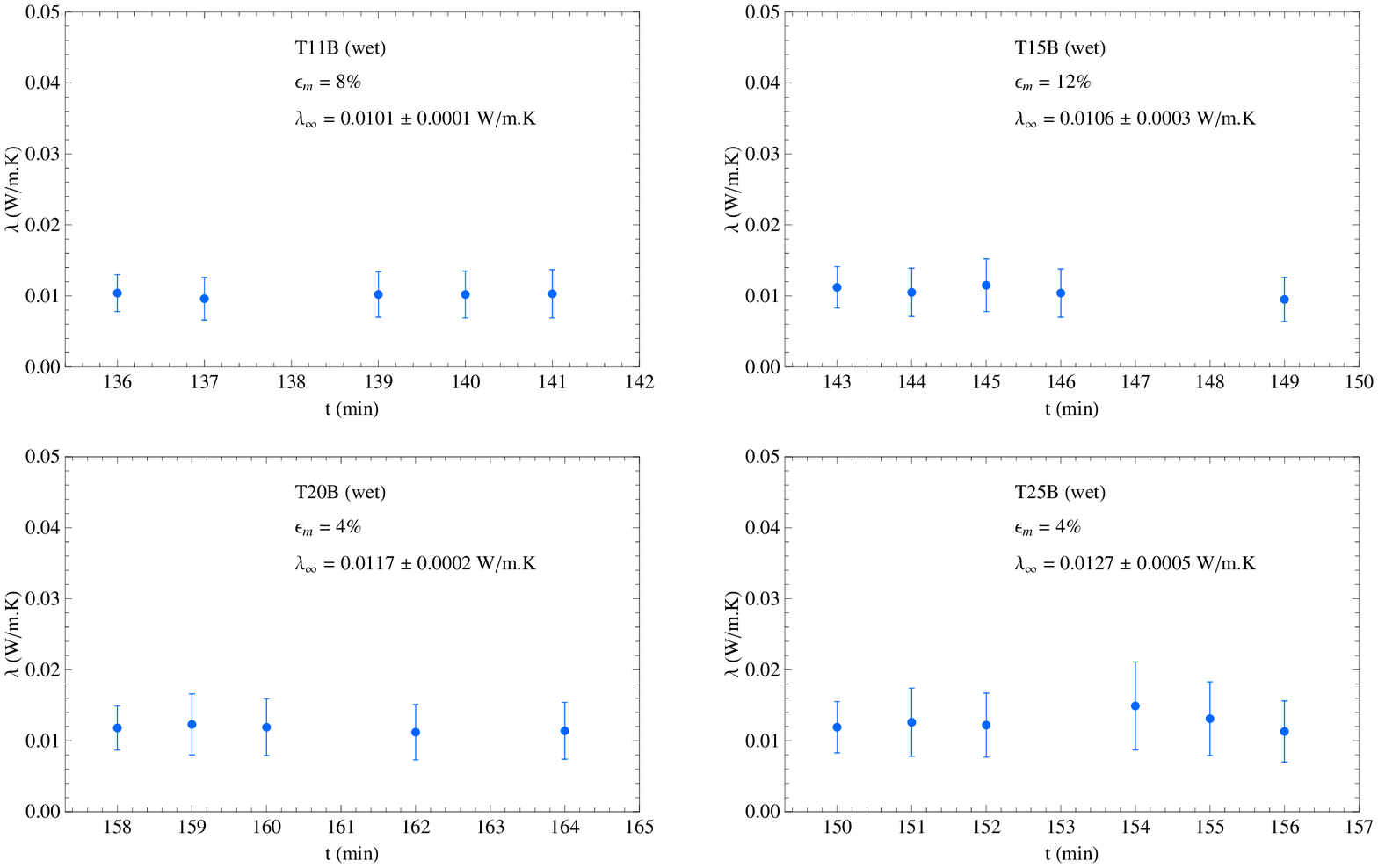}
\end{tabular}
\caption{Thermal conductivity for type $A$,$B$ fabrics soaked with water.}
\label{fig6}
\linea
\end{figure}

\noindent The next step we undertook has been to extend our measurement protocol to study the behavior of the cotton samples with respect to water retaining. In this case, of course, we have to take a proper account of water evaporation, and the following method has been applied in order to obtain consistent experimental data for a given water concentration, here measured indirectly from the mass of the wet fabric. The basic idea is to measure the mass of the dry fabric before any thermal measurements, and then to measure the mass of the wet fabric shortly before and after the collecting of temperature data, in order to obtain a given average concentration to be directly related to the actual thermal measurement. In the present case, we have chosen to collect data from each sample until water concentration was reduced to about a quarter of the initial one, that is when
\begin{equation}
M_{\rm{wet}} - M_{\rm{dry}} = \frac{1}{4} \left( M_{\rm{wet, \,initial}} - M_{\rm dry} \right) . 
\end{equation}
The measured values of the thermal conductivity for the wet sample were, then, obtained as above, and plotted in Fig.s\,\ref{fig6} for a given value of the dimensionless mass concentration of water that the given sample had initially retained,
\begin{equation}
\varepsilon_m = \frac{M_{\rm{wet}} - M_{\rm{dry}}}{M_{\rm{dry}}} \, .
\end{equation}

The experimental data we obtained (see Table \ref{tab3}) confirm an important physical property: when the air prevails within the fabric, its conductivity is lower, so that the transpirability is higher and a \textit{warm} sensation is induced. Instead, a non-negligible water concentration leads to higher values of the conductivity, so that the transpiring capacity of the fabric is reduced and, therefore, the present moisture makes the fabric ``sticky'', providing a \textit{cold} sensation \cite{Norma}.

\begin{table}%[!h]
\centering
\begin{tabular}{|c|c|c|c|c|}
\hline \textbf{Fabric} & $\lambda_{\rm Dry}$ \footnotesize{($10^{-4}$ W/m K)}
& $\lambda_{\rm Wet}$ \footnotesize{($10^{-4}$ W/m K)} & $\varepsilon_m$ \\
\hline T11A &    $85 \pm 7$ & $119   \pm 5$    & 4\% \\
\hline T15A &	$76 \pm 7$ & $111  \pm 3$    & 2\% \\
\hline T20A &	$60 \pm 10$ & $147  \pm 5$    & 4\% \\
\hline T25A &	$96 \pm 8$ & $111   \pm 7$    & 4\% \\
\hline T11B &	$92 \pm 3$ & $101 \pm 1$ & 8\% \\
\hline T15B &	$96 \pm 2$ & $106 \pm 3$    & 12\%\\
\hline T20B &	$140 \pm 4$ & $117 \pm 2$ & 4\% \\
\hline T25B &	$120 \pm 3$ & $127    \pm 5$    & 4\% \\
\hline
\end{tabular}
\caption{Thermal conductivity of different cotton samples for dry and wet fabric (regime values).}
\label{tab3}
\end{table}

From a physical point of view, water fluence through the fabric depends mainly on the nature of the micro-pores present in the tissue, whose thermal behavior is, on its turn, directly influenced by water: from the experimental data 
in Fig.s\,\ref{fig6} it is clear that the regime condition is achieved sooner in time. As a consequence, the exponential best fit, adopted in previous measurements, is now replaced by an horizontal straight line: that is, thermal conductivity reaches directly a constant value (on the other hand, by using again an exponential fit we get identical results). The shorter time required for approaching the regime condition is a clear indication that water within the fabric modifies the structural composition of the fabric itself. The convective action of water in the heat propagation mechanism makes the system to get thermal homogeneity much early in time, while the specific heat turns out steadier. Indeed, the specific heat of the wet fabric is different from that of the dry one and also changes with temperature, thus directly influencing the conductivity. As a consequence, the kinetic of global thermalization becomes flatter and reaches earlier its regime value.

\section{Conclusions}

\noindent The measurement protocol presented here has been tested by means of very long calibration procedure, aimed at acquiring a proper knowledge about the physical behavior of the relevant components of our experimental apparatus. In particular, we have built up appropriate calibration models for the measurements of thermal conductivity of fabrics, and applied standard regression methods to the statistical analysis of the calibration operation. Our measurements on cotton fabrics have provided useful information about their behavior in different physical conditions: wet and dry fabrics undergo a different thermalization kinetics and have different conductivities, leading to different comfort features. Best fit curves for any relevant thermal quantity have, moreover, been obtained from the measurements on dry fabrics.

Differently from other devices, the {\it ThermoTex} protocol associated with the {\it Thecotex} machinery applies rigorous physical procedures, and then leads to a closer agreement with the international scientific standards. The calibration patterns for the heat source have been obtained for different operating voltages and, therefore, for different operating temperatures: the typical trend exhibited by the data taken in practically any given condition (see Fig.\,\ref{fig2}a) allows a thorough enlargement of the range of temperatures at which measurements and tests can be performed.

The measured values of thermal conductivity show a sensitivity to large temperature variations, as expected for many insulating materials.

Actually, the ThermoTex protocol applies to many different types of materials (not only fabrics), and to very different physical conditions. Our present analysis of the different dry/wet fabric behavior has, in particular, confirmed experimentally how the physical properties of water, having a higher conductivity, influence the thermalization kinetics of the cotton sample, the wet samples reaching the regime conductivity condition earlier in time.

The detailed study of the physical behavior of our experimental setup allows further improvements of it, including a new electronic system driving the automatic collection of experimental data and the automatic calibration of the power transmitted by the hot plate in the course of time. In particular, our future aim will be to extend the present measurements to a larger number of fabrics and other materials in different configurations, analyzed -- again -- with the rigorous approach described above. The final step of the present research program will be a phenomenological theory of heat transfer through the fabrics, able to predict characteristic comfort properties to meet the market demand. The predictions of such a theory about relevant textile thermal properties could, then, be tested, and novel textile products, suitable for different applications, could finally be manufactured. 

\begin{acknowledgements}
\noindent We are indebted with C. Baldassarri, S. Bonfanti, I. Chiodo and C. Colleoni for their invaluable help during the experimental work presented here.
\end{acknowledgements}

%\cleardoublepage

\end{document}